\documentclass{elsart}
\newcommand{\be}{\begin{eqnarray}}
\newcommand{\ee}{\end{eqnarray}}
\usepackage{amsmath}
\usepackage{amssymb}
\usepackage{epsfig}

\newcommand{\dis}{\displaystyle}
\newcommand{\mn}{M_N}
\newcommand{\mpi}{m_{\pi}}

\begin{document}
\hfill{\small FZJ--IKP(TH)--2004--01, HISKP-TH-04/03}
\begin{frontmatter}
\title{The role of the nucleon recoil in low--energy meson--nucleus reactions}
 
\author{V. Baru$^{1,2}$,  C.~Hanhart$^1$, A.E.~Kudryavtsev$^2$, and U.-G. Mei\ss ner$^{1,3}$}

{\small $^1$Institut f\"{u}r Kernphysik, Forschungszentrum J\"{u}lich GmbH,}\\ 
{\small D--52425 J\"{u}lich, Germany} \\
{\small $^2$Institute of Theoretical and Experimental Physics,}\\
{\small 117259, B. Cheremushkinskaya 25, Moscow, Russia}\\
{\small $^3$Helmholtz-Institut f\"{u}r Strahlen- und Kernphysik (Theorie), 
Universit\"at Bonn}\\ 
{\small Nu{\ss}allee 14-16, D--53115 Bonn, Germany} 
 
\begin{abstract}
The role of the nucleon recoil corrections in low--energy meson--nucleus
interactions is examined. We demonstrate explicitly when calculations within
the static approximation are justified and when the recoil terms need to be kept
explictly in the propagators, depending on whether the S-wave 
two--nucleon intermediate
state is Pauli blocked or not, while the meson is in flight. 
While the effect is studied in detail for $\pi d$ scattering only, 
a large class of other reactions is discussed for which the findings are
relevant as well.
\end{abstract}

\end{frontmatter}

\section{Introduction}

Low--energy meson--nucleon reactions are of great theoretical interest for
they are one of the best tools to deepen our understanding of the nuclear
many-body problem. The production and scattering of the lightest
member of the Goldstone nonet, i.e. the pion, on nuclei is the subject of
special experimental and theoretical interest since they allow to test
predictions of chiral perturbation theory and---within this scheme---quantify
the effect of isospin violation in the strong $\pi N$ interaction.  In
addition, pion-nucleus reactions
 can be used to get information
on the elementary pion--neutron interactions. 
Detailed knowledge of the latter is important to fix the isoscalar
pion--nucleon scattering length.  
The calculation of the production and scattering
processes for heavier mesons on nuclei is more difficult but
not less interesting. For instance, the reactions involving the $\eta$
meson can be used to explore the possibility of the formation of the
$\eta$-nucleus bound state, whereas the $Kd$-scattering allows to
extend our knowledge about the strange sector.

It is well known that rescattering effects, where the intermediate meson being
scattered on one nucleon of the nucleus  then rescatters on another one, 
are potentially large and need to be evaluated in a controlled way in
order to, for example, extract information on the the neutron amplitudes.
 How to do this within an
effective field theory Weinberg described in one of his classic papers
\cite{wein}. In this paper nucleons are treated rigorously as static, as long
as the diagram is two--nucleon irreducible, leading
to largely simplified nuclear matrix elements. This framework
has been subsequently extended to higher orders and consistent
wave functions based on chiral nuclear effective field theory have
been used, see \cite{BBLM,BBEMP}.

In this note we 
investigate under what circumstances the static approximation is
justified and also identify reactions where the
recoil corrections are to play a significant role. 
On the example of the calculation of the $\pi d$ scattering length  we
show that 
cancellations amongst different one--body and two--body amplitudes have to
occur in order to make the Pauli principle also hold for two--nucleon states
while there is a pion in flight simultaneously. Similar arguments were
recently presented by Rekalo et al. \cite{rekalo} for neutral pion photo- and
electroproduction on the deuteron at low energies. However, these authors
argue that in case of a Pauli forbidden intermediate state the pion
rescattering contribution has to be canceled completely in order to allow the
nucleons to obey the Pauli principle.  In this paper we critically reexamine
this claim. Especially we find that it is not the full rescattering amplitude
that vanishes in case of the Pauli forbidden S-wave intermediate
two--nucleon state, but
only the part stemming from the nucleon recoils, leaving the static exchange
as a good approximation to the full result.  Note that the interference pattern
discussed was already observed
numerically 
in the phenomenological approach of Refs. \cite{BK97,BKT}.
In addition, we will show that, at least for $\pi d$ scattering near
threshold, for cases where the S-wave intermediate two--nucleon state is Pauli
allowed, the net effect of the (then important) recoil corrections is that
the corresponding rescattering contributions (i.e. the static term +
the corrections occuring due to the finite nucleon mass) are numerically irrelevant.

Historically the presence of sizable cancellation in
calculation for the $\pi d$ scattering length was  raised first in
the papers by Kolybasov et al. \cite{Kolyb} and independently by
F\"aldt \cite{Faeldt} 
where it was claimed that  the naive static term is a good
approximation for rescattering effects
(see also discussion in Ref. \cite{Weise}). However, the effect in
Ref. \cite{Faeldt}  was traced to
a quite unnatural numerical cancellation of the diagrams shown in Fig. 1 b)
and c) with the corresponding diagrams
where the intermediate nucleons are rescattered off each other.
Moreover in  Ref. \cite{Faeldt}  rescattering effects are approximated 
by the static term for both types of the S-wave $\pi N$-potential,
i.e. for isoscalar and isovector. This is indeed correct for the isovector
$\pi N$-interaction but not for the isoscalar one as we will prove  below. 
Due to the smallness of the isoscalar $\pi N$-scattering length the
wrong interpretation of rescattering effects in this case does not
affect the full result of the calculation of the $\pi d$-scattering
length, however, for other processes where this term is not small one
would get the incorrect result.

Note that arguments based on the Pauli
principle for the intermediate nucleons are very general and therefore
can be applied for different processes with any intermediate meson
exchanges. Several examples will be discussed in sec. \ref{results}.

\section{Pauli principle and the $\pi d$ scattering length}

To be concrete, this discussion will be carried out for the reaction $\pi d\to
\pi d$ and we will restrict ourselves to a rather simplified $\pi N$
interaction that, however, allows us to address all relevant issues.

\begin{figure}[t!]
\begin{center}
\epsfig{file=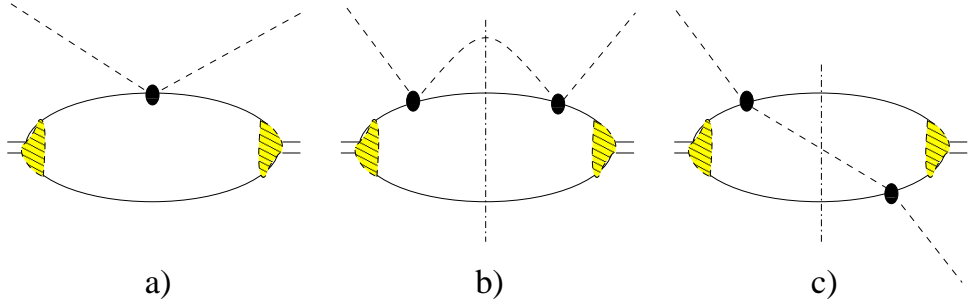, height=4cm, angle=0}
\caption{Typical Feynman diagrams for $\pi d$ scattering; shown are one--body terms
  $\bigl ($diagram a) and b), as well as the corresponding
 rescattering   contribution c)  $\bigr)$. Solid
  black dots stand for the $\pi N$ interaction, whereas the hatched area shows
  the deuteron wave function. Crossed terms (where the external pion lines are
 interchanged) are not shown explicitly.}
\label{dia}
\end{center}
\end{figure}

In Fig. \ref{dia} we show typical single nucleon 
diagrams a) and b) as well as the 
corresponding two--nucleon  contribution c).  We will concentrate on the
isolation of the one--body (isoscalar) amplitude (as it would be also
measured in a $\pi N$ scattering experiment or extracted from a partial-wave
analysis) from that related to nuclear effects.

 In general, two--nucleon states in any subsystem have to obey the
Pauli principle. Thus, when looking at diagram c), where the relevant
intermediate state is marked by a perpendicular line, the Pauli principle
demands the two nucleons to show a particular behavior under their
exchange. However, if we would exchange the two nucleons in the intermediate
state of diagram c) we come automatically to diagram b). Thus, we should expect
some destructive interference between b) and c) if the intermediate state is forbidden
by the Pauli principle and a constructive one if the intermediate state is
allowed. 

The interference pattern depends on the particular
structure of the $\pi N\to \pi N$ transition potential. 
At the threshold it may be written as (the factor 1/$\sqrt{2}$ was introduced
for convenience since this way we will get results symmetric in $g_+$ and $g_-$) 
\begin{equation}
V^{ba}_{\pi N}=\delta^{ba}g_++\frac i{\sqrt{2}}\epsilon^{abc}\tau^cg_- \ ,
\label{pinpot}
\end{equation}
where the strength of the isoscalar (isovector) interaction is denoted by
$g_+ \, (g_-)$ and $a$\, ($b$) is the isospin index of the initial (final) pion.
For our study we assume both $g_+$ and $g_-$ as constant.
The conclusions do not change by this simplification, but the formulas
simplify largely.
As was shown by Weinberg \cite{Weinberg66}, Tomozawa \cite{Tomozawa66} and
others,
$(g_+/g_-) \sim (m_\pi/\Lambda)$, where $\Lambda$ denotes the typical hadronic
scale of order 1 GeV and $m_\pi$ is the pion mass. Thus, in $\pi d$ scattering
the isoscalar rescattering contribution is largely suppressed.
However, for illustration we will keep both $g_+$ and
$g_-$ in the calculation. 

Lets us now analyze more closely the structure of the matrix element
corresponding to diagram \ref{dia}c). The deuteron is an isoscalar. Thus, when
a scalar (vector) operator in isospin space 
 operates on this state,
the resulting two--nucleon state is in an isoscalar (isovector) state. 
At threshold the $\pi N$ interaction is spin and momentum--independent and therefore the 
two--nucleon state after the $\pi N$ interaction is in both scenarios  in a spin
triplet $S$--wave state. The Pauli principle allows this only for isoscalars
and thus, if the $\pi N$ interaction is given by the $g_+$ term, the
contributions of diagrams c) and b) to the $\pi d$ scattering length 
due to $NN$ recoil should
interfere constructively and, if it is given by the $g_-$ term, they should
interfere destructively. 

Within our simplified model the $\pi d$ scattering length is given by 
\begin{equation}
a=a_a+a_b^++a_b^-+a_c^++a_c^-,
\label{apid}
\end{equation}
where the superscripts denote the isospin structure of the $\pi
N$-interaction and the subscripts refer to the diagrams as shown in
Fig. \ref{dia}.
Since the deuteron is an isoscalar state, only $g_+$ contributes to
diagram a), and we get 
\begin{equation}
a_a=\frac{2g_+}{4\pi \left (1+\ m_{\pi}/({2M_N}) \right)}.\ 
\label{pot}
\end{equation}
We also give the expressions for  diagram b)
\begin{equation}
a_b^\pm= g_\pm^2\,\xi\int
d^3pd^3q\Psi(\vec p)^\dagger \frac{1}{\vec q\, ^2+\rho}\Psi(\vec p) \ 
\label{1bloop}
\end{equation}
and for diagram c) 
\begin{equation}
a_c^\pm=\pm g_\pm^2\,\xi \int
d^3pd^3q\Psi(\vec p - \vec q)^\dagger \frac{1}{\vec q\, ^2+\rho}\Psi(\vec p) \ ,
\label{rescat}
\end{equation}
where $\Psi$ denotes the deuteron wave function 
and  p (q) is the relative nucleon momentum before the $\pi N$ interaction (the 
intemediate pion  momentum).
Furthermore,  $\xi=[16\pi^4 \left (1+
  m_{\pi}/(2M_N) \right )]^{-1}$ , $\rho = \omega(2\epsilon +({\vec p\,
  }^2+(\vec p - \vec q)^2)/M_N)$, with $\epsilon$ and $\omega$ for the
deuteron binding energy and the pion energy, respectively, and $M_N$ is the
nucleon mass.
To come to these expressions 
terms of order $\rho^2$ ($\rho$) were dropped in the denominator (numerator),
since they lead only to the complication of the formulae but are
irrelevant for this study. The piece linear in $\rho$ that appears in the
denominators is the recoil correction of interest. In
Ref. \cite{wein} as well as many multiple scattering formalisms (see
Ref. \cite{Weise}
and references therein)
the structure given in Eq. (\ref{rescat}) with $\rho=0$ is given for the
rescattering contribution.
However, when straightforwardly evaluated the inclusion of $\rho$ 
decreases the value of $a_c$ approximately by a 
factor of 2 \cite{BK97} (see also Ref. \cite{KK} where the recoil corrections 
were estimated for the first time and found to be important).
The reason for this relatively large effect of the
recoil corrections on the amplitude is the proximity of a three--body
singularity: the fact that for $\epsilon =0$ the intermediate
states can become real leads to a contribution non--analytic in the nucleon
mass that can not be simply dropped.

The goal of our study is two-fold: we want to split the $\pi d$ amplitude into
its contribution from the $\pi N$ amplitude and the nuclear corrections---the
so--called rescattering contributions---and secondly we want to understand the
role of the nucleon recoil in the latter.  The former issue was previously
addressed within chiral perturbation theory in Refs.  \cite{wein,BBLM,BBEMP},
but this separation is much more transparent in our simplified approach and
thus allows to discuss the various contributions more directly.  Thus, after
adding and subtracting appropriate terms, we decompose
the expression for the $\pi d$ scattering length
 in the following way
\begin{equation}
a=a^{(1-body)}+a^{(static)}_{LO}+a^{(recoil)}+a^{(static)}_{NLO} \ .
\end{equation}
Here the one--body piece is given by
\begin{equation}
a^{(1-body)}=\frac{2g_+}{4\pi \left (1+\ m_{\pi}/({2M_N})
  \right)}+(g_+^2+g_-^2)\,\xi \int
d^3q\frac{1}{\vec q\, ^2 +\tilde \rho} \ ,  
\label{1bodydef}
\end{equation}
where $\tilde \rho = \omega \vec q\, ^2/M_N$ and a particular
regularization prescription is to be given to render the 
loop integral finite (see e.g. Ref. \cite{ulfs}
). Note that within our model the full expression given in
Eq.~(\ref{1bodydef}) is the expression for the isoscalar $\pi N$ scattering
length. It is therefore this piece that we need to isolate, if we want to
extract single nucleon amplitudes from nuclear reactions. 
The rescattering contribution (or nuclear corrections) we split
without any further approximations into  the leading order static 
piece given in Ref. \cite{wein}
\begin{equation}
a^{(static)}_{LO}=(g_+^2-g_-^2)I_0
\label{stat}
\end{equation}
with
\begin{equation}
I_0=\xi \int
d^3pd^3q\Psi(\vec p - \vec q)^\dagger \frac{1}{\vec q\, ^2}\Psi(\vec p) \ ,
\end{equation}
and the corrections occuring due to the finite nucleon mass
which are again separated on two parts, namely the
3--body singularity correction (or the recoil term) which is not
analytic in $\omega_{\pi}/M_N$ and the term
$a^{(static)}_{NLO}$ which is regular in
$\omega_{\pi}/M_N$. 
 These corrections can be expressed as
\begin{eqnarray}
{a}^{(recoil)}&=&{g_+^2}I_++{g_-^2}I_-~,
\label{rescat2}
\end{eqnarray}
with
\begin{eqnarray}
\hspace{-0.5cm} I_\pm=&\frac\xi 2\int
d^3pd^3q|\Psi(\vec p)\pm\Psi(\vec p-\vec q)|^2 \left(\frac{1}{\vec q\, ^2+
  \rho}-\frac{1}{\vec q\, ^2+  \tilde \rho }\right)~,
\label{idef}
\end{eqnarray}
where we used the symmetry of $I_\pm$ to replace $\Psi(\vec p)$ by 
$\dis\frac{1}{2}(\Psi(\vec p)\pm\Psi(\vec p-\vec q))$ 
and
\begin{eqnarray}
a^{(static)}_{NLO}=(g_+^2-g_-^2)\Delta I~,
\label{NLO}
\end{eqnarray}
with
\begin{eqnarray}
\Delta I&=&\xi\int
d^3pd^3q\Psi(\vec p-\vec q)^\dagger \left(\frac{1}{\vec q\, ^2+
 \tilde \rho}-\frac{1}{\vec q\, ^2}\right)\Psi(p) \simeq
-\frac{m_\pi}{M_N} I_0 \ .
\label{delI}
\end{eqnarray}
Note that $a^{(static)}$, $a^{(recoil)}$
and $a^{(static)}_{NLO}$ are finite.


We numerically evaluated $I_\pm$ and $I_0$ using the deuteron wave functions from the
Bonn potential \cite{bonn} and found 
$$
I_+= -0.88 \ , \ \mbox{and} \qquad I_-= -0.19~,
$$
both given in units of $I_0$. We checked that the numbers do not depend on
the type of the deuteron wave function used. 
The results  reflect the interference
pattern discussed above, i.e. that the 3--body correction due to the
nucleon recoil is much smaller in the case when the S-wave $NN$
intermediate state is forbidden by the Pauli principle compared to
that when it is allowed ($I_-\!\ll\! I_+$). 
To make it even more clear that the interference
pattern in $I_\pm$ indeed reflects what is demanded by the Pauli principle let
us express the integrals explicitly in terms of the two--nucleon relative
momentum, $\vec p\, '=\vec p-\vec q/2$, for the $NN$ intermediate state. Then
the first term of the integrant in Eq. (\ref{idef}) reads
$$
\Psi\left(\vec p \, '+\frac12\vec q\, \right)
\pm\Psi\left(\vec p \, '-\frac12\vec q\, \right).
$$
For the upper (lower) sign this combination is symmetric (antisymmetric)
under the transformation\footnote{For the deuteron D-wave this is correct
  since $Y_2^m(-\hat p)=Y_2^m(\hat p)$} $\vec p\,'\to-\vec p\,'$.  Thus, in
case of $I_+$ ($I_-$) this term projects on the Pauli allowed
 even (odd) two--nucleon spin triplet states in
the $NN\pi$ intermediate state---a statement that holds for both the imaginary
part as well as the real part!

Note also that the contributions of diagrams 1 b) and c) to the
imaginary part of the $\pi d$ scattering length (e.g. due to charge
exchange process) can appear only from the recoil term, i.e. from
$I_+$ ($I_-$). In case of the calculation of the imaginary part of
$I_+$ ($I_-$) one has to integrate over the momenta in the narrow
region restricted by the 3--body phase space, i.e. over the region
where the contribution of the P-wave $NN$ state is small. 
This  leads to a rather small value of ${\rm Im} \, I_-$.


We stress  that $I_+$ is large and negative. Therefore when adding the terms
proportional to $g_+^2$ in ${a}^{(recoil)}$ and $a^{(static)}$, they largely
cancel. On the other hand, $I_-$ is much smaller than $I_0$ and therefore the nuclear
contributions to the $\pi d$ scattering amplitude proportional to $g_-^2$ are
basically given by $a^{(static)}$.
 Thus, we find that if the S-wave two--nucleon intermediate state
that occurs while the pion is in flight is allowed by the Pauli principle,
the net effect of the rescattering contribution is quite small. On the
other hand, if the S-wave two--nucleon intermediate state
 is Pauli forbidden, the rescattering effects are large, however, can
 be well  approximated by a static exchange. The corrections to this
 are found to be of order $15-20\%$ of the static term.
However, one should also not forget about the NLO correction to the
static term. This correction is not related with the Pauli principle
and it is regular in $\omega/M_N$. Moreover, in distinction from the
3--body recoil correction, which is only weakly dependent on the
mass of the exchanged meson, the NLO correction to the static term
linearly depends 
on the mass. It turned out that for the $\pi d$ scattering the integral
$\Delta I$ is equal to $-0.15$ (in the same units), thus making the conclusion
about the role of rescattering effects presented above even stronger. Indeed,
this term interferes destructively with $I_-$ (constructively with $I_+$)
resulting in the total correction to the static term (Eq. (\ref{stat}))
to be equal to 
$$\! a^{(recoil)}+a^{(static)}_{NLO}=g_+^2 (I_++\Delta
I)-g_-^2(-I_-+\Delta I)
=(-1.03 g_+^2 -0.04 g_-^2)I_0.$$
Thus, the total rescattering contribution is $(-0.03 g_+^2 -1.04
g_-^2)I_0$, i.e. it is basically negligible for the isoscalar part of
the $\pi N$ interaction and it
almost equals to the static term for the isovector one. 
However, this particular cancellation
seems to be specific to $\pi d$ scattering and the result might be different
for exchange mesons heavier than the pion.

As one might expect intuitively, Pauli principle arguments do not apply to the
static piece, for it describes the instantanous exchange of a pion that does
not allow for any two--nucleon intermediate state. However, one comment is
necessary at this place. 
It might be that
there is still a potentially sizable cancellation
between the terms proportional to $g_-^2$ in $a^{(1-body)}$ and
$a^{(static)}_{LO}$, simply because these terms are of different signs. 
However, the goal of the present study is to properly isolate the
one--body piece from the rescattering piece
---this is
the precondition to, for example, extract neutron amplitudes from deuteron
observables. The isoscalar $\pi N$ scattering length, however, is
proportional to the full expression given in  Eq. (\ref{1bodydef}) and not
just its first term.

\section{Discussion}
\label{results}

As mentioned above, in reality the isovector $\pi N$ interaction is
significantly stronger than the isoscalar one in the near threshold
regime. Therefore, for this case  we may drop all terms proportional to
$g_+^2$ and find, that indeed the static approximation is justified to
calculate the rescattering corrections.

The same kind of selection rules as discussed here for $\pi d$ scattering
obviously also apply to other meson--nucleus scattering reactions.  
The first example we will discuss here is coherent $\pi^0$ photo- and
electroproduction on the deuteron. 
For this reaction the reasoning is basically identical. Although the 
leading $\gamma \pi \bar N N$ vertex of Kroll--Rudermann 
type\footnote{This refers, of course,
to the rescattering corrections since the Kroll-Rudermann vertex only leads 
to charged pion production.} contains both a spin and
an isospin operator and thus can lead to the $^1S_0$ intermediate $NN$ state, the rescattering
vertices again are of the type given in Eq. (\ref{pinpot}).
Based on this observation in Ref. \cite{rekalo} it was
concluded that the Pauli principle leads to a vanishing of all rescattering
contributions. However, given the discussion at the end of the previous
section, we now understand that this conclusion is based on an improper
separation of one--body and rescattering terms and indeed a static pion
exchange is a good approximation for the leading rescattering contribution.
Therefore the quite elaborate calculations of
neutral pion photo- and electroproduction based on chiral perturbation theory
with static nucleons
\cite{BBLMvK,KBM} give accurate results.

However, the picture changes when we look at incoherent pion production
reactions like $\gamma d\to \pi^+nn$ and for this reaction large effects from
the recoils are to be expected. In this context it is interesting to note
that the existing data for this reaction are well described by the one--body
terms alone, however, the inclusion of rescattering contributions, estimated
in the static approximation, lead to a visible deviation of the
calculation from the data (see the review \cite{laget} and
references therein). A possible solution to this is that the recoil
corrections strongly  diminish the rescattering piece, as discussed
above, leaving the one--body piece as a good approximation to the full result.
We will check this conjecture by an explicit calculation in a subsequent
publication.

Another case of interest is $\eta$--nucleus scattering. For example in a
recent analysis of the world data set on $pd\to \eta ^3$He indications were
found that the imaginary part of the $\eta ^3$He scattering length might be
significantly smaller than three times the imaginary part of the elementary
$\eta N$ amplitude \cite{etahe3}, contrary to what is expected from many
multiple scattering approaches. Given the findings of above, this should no
longer come as a surprise: the $^3$He wave function is dominated by a
quasi--deuteron together with an additional proton in an $s$--wave
\cite{hepara}. In addition, the dominant piece of the $\eta N\to \pi N$
transition amplitude close to the $\eta$--threshold is spin and
momentum--independent, for it is driven by the resonance $S_{11}(1535)$, but
is isospin--dependent.  Consequently, all the reasoning given above applies
and to a good approximation the imarinary parts that originate from the
one--body amplitudes have to be cancelled by those from the rescattering and
therefore there is no connection between the imaginary part of the $\eta ^3$He
scattering length and that for $\eta N$ scattering.

\section{Summary}

In this letter, we have studied the role of recoil corrections in low--energy
meson--nucleus reactions. In a simplified model of $\pi d$ scattering, we
have shown how to separate the one--body contribution, that embodies the
pertient information on the elementary pion-nucleon amplitude, from the 
rescattering (two-body) corrections, paying particular attention to the
constraints from the Pauli principle.
Our results are opposite to those of Ref. \cite{rekalo}, where it was
argued that rescattering contributions are negligible when Pauli forbidden
intermediate states occur whereas they are large for Pauli allowed
states. The difference was traced to an improper separation of
one--body and nuclear contributions in Ref. \cite{rekalo}.  Our
findings indicate, that, if  the S-wave $NN$ intermediate state is
Pauli forbidden, the static
meson exchange is a good approximation to the full amplitude, whereas in case
of the Pauli allowed S-wave intermediate state the (then significant) recoil corrections
strongly suppress the nuclear corrections.
In this paper the consequences of these insights for other reactions were discussed.

\vfill

The authors thank A. M. Gasparyan  for many interesting discussions.
A.E.K. also acknowledges  V.G. Ksenzov for useful discussions. 
This work was partly supported by RFBR grant N$^0$~02-02-16465 and by 
02-02-04001 DFG-RFBR grant 436 RUS 113/652.

%
%

\end{document}